# Electrostatic shielding effects and binding energy shifts of molybdenum chalcogenides

Yaorui Tan, Maolin Bo*

[a] Y. Tan, M. Bo

Key Laboratory of Extraordinary Bond Engineering and Advanced Materials Technology (EBEAM) of Chongqing

Yangtze Normal University

Chongqing 408100, China

*Author to whom correspondence should be addressed: bmlwd@yznu.edu.cn (Maolin Bo)

**Abstract:** This study examines the electronic structure and bond properties of $MoS_2$, $MoSe_2$, and $MoTe_2$. Density functional theory (DFT) and the binding energy and bond-charge (BBC) model are applied to calculate their electronic structure, binding energy shift, and bond properties. It is found that electrostatic shielding by electron exchange is the main cause of density fluctuation. Finally, this study establishes a method for calculating the state density with Green's function with energy level shift and provides new methods and ideas for the further study of the binding energy, bond states, and electronic properties of nanomaterials.

## Introduction

In the last two decades, two-dimensional (2D)-layered transition metal dihalides (TMDs), including $MoS_2$[1-5], $MoSe_2$[6, 7 8], and $MoTe_2$[9-11] materials, have spurred breakthroughs in light-emitting diodes and in photodetection and other optical devices. Moreover, the different phase 2D structures of $MoS_2$, combined with its large specific surface area advantage, have found used as catalysts or carriers in traditional multiphase thermal catalysis, photocatalysis, and electrocatalysis, and $MoS_2$ nanosheets have been applied as catalysts of the hydrogen evolution reaction[12]. As the optical band gap range of $MoSe_2$ is 1.2–1.3 eV, it can absorb visible and near-infrared light, making it widely used in the fields of field-effect transistors[13] and optoelectronics. Similarly, $MoTe_2$ is a typical topological quantum material[14, 15] of the $MX_2$ family of layered transition metal chalcogenides. Therefore, TMDs materials have been extensively studied and reported recently.

X-ray photoelectron spectroscopy (XPS) is a powerful tool for studying the surface properties of materials, including their binding energies (BEs). The BE of the same element is different in different compounds, giving rise to the phenomenon of energy shift, which reflects the electronic structure and bond properties of the material. Accordingly, we studied the electronic structure and bond properties of $MoS_2$, $MoSe_2$, and $MoTe_2$ through their BE shifts.

In this paper, we use DFT and the BBC model to calculate the electronic structure and bond properties of $MoS_2$, $MoSe_2$, and $MoTe_2$[16], finding that electrostatic shielding by electron exchange is the main cause of density fluctuation. Combined with band theory and Green's function, the relationship between energy shift and lattice density is established, and the overlapping integral and three-dimensional bond-electron band are obtained by lattice density calculation. It provides new methods and ideas for the further study of the atomic bonding and electronic properties of nanomaterials.

## Methods

### DFT calculations

The atomic bonding, electronic properties and structural relaxation of $MoS_2$, $MoSe_2$, and $MoTe_2$ materials were investigated with the Cambridge Sequential Total Energy Package (CASTEP), which uses total energy with a plane-wave pseudopotential. This analysis focused on the bonding, structure, energetics, and electronic properties of $MoS_2$, $MoSe_2$, and $MoTe_2$ materials. We used the Perdew–Burke–Ernzerhof (PBE)[17] functional to describe their electron exchange and correlation potentials. We used the TS scheme for DFT-D dispersion correction to consider the long-range vdW interaction. The cutoff energy, lattice parameters, and k-point grids are as shown in Tables 1 and 2. In the calculations, the vacuum space was approximately 14 Å. the energy converged was $1\times10^{-6}$ ,the force on each atom converged to < 0.01 eV/ Å.

### BBC model

The BBC model is quantified bonds by BE shift and deformation charge density. The binding-energy (BB) model can get the relationship between the energy shift and Hamiltonian. The bond-charge (BC) model use deformation charge density to calculate atomic bonding states.

For BC model, considering the effect of external fields on BE shifts:

$$H' = \xi(0)\sum_{l}\hat{n}_l - t\sum_{l}\sum_{\rho}C_l^{+}C_{l+\rho} = \left(E_n^a - \gamma A_n\right)\sum_{l}\hat{n}_l - \gamma t\sum_{l}\sum_{\rho}C_l^{+}C_{l+\rho}$$

(1)

$\hat{n}_l$ represents the electron number operator in the wanier representation. So $\sum_{l}\hat{n}_l = C_l^{+}C_l$ can be obtained.

$A_n$ represents the energy shift, $E_n^a$ is the atomic energy level, and *t* is called the overlapping integral.

$$\begin{cases} A_n = -\int a_n^*\left(\vec{r}-\vec{l}\right)\left[V\left(\vec{r}\right)-v_a\left(\vec{r}-\vec{l}\right)\right]a_n\left(\vec{r}-\vec{l}\right)\mathrm{d}r \\ E_n^a = \int a_n^*\left(\vec{r}-\vec{l}\right)\left[-\dfrac{\hbar^2}{2m}\nabla^2 + v_a\left(\vec{r}-\vec{l}\right)\right]a_n\left(\vec{r}-\vec{l}\right)\mathrm{d}r \\ t = \sum_{l,l'}C_l^{+}C_{l'}\int a^*\left(\vec{r}-\vec{l}\right)\left[V\left(\vec{r}\right)-v_a\left(\vec{r}-\vec{l}\right)\right]a\left(\vec{r}-\vec{l'}\right)\mathrm{d}r \end{cases}$$

(2)

$\xi(0)$ is the local orbital electron energy. Therefore, there exists $\xi(0)=\left(E_n^a - A_n\right)$ . $E_n^a$ caused by the potential of *N*-1 atoms outside the lattice point *l* in the lattice. $V(r)$ is the potential and $v_a\left(\vec{r}-\vec{l}\right)$ is the atomic potential.

The vth orbital electronic energy shift of atoms:

$$\Delta E_v(B) = A_n\sum_{l}\hat{n}_l,\quad \Delta E_v(x) = \gamma A_n\sum_{l}\hat{n}_l$$

(3)

$\Delta E_v(B)$ is represents the energy shift of an atom in an ideal bulk. $\delta\gamma = \gamma - 1$ is relative BE ratio and B indicates bulk atoms. The effective positive charge of the ion is $Z' = Z - \mu_v$, considering the charge shielding effect $\lambda_v$, where Z is the nuclear charge.

Further based on $\Psi(\vec{r}) = \sum_l C_l a(\vec{r} - \vec{l})$ , $V_{cry}(\vec{r}_i - \vec{l}_j) = V(\vec{r}) - v_a(\vec{r} - \vec{l})$ and $\gamma V_{cry}(\vec{r}_i - \vec{l}_j) = \gamma \sum_{i,j,l_j \neq 0} \frac{1}{4\pi\varepsilon_0} \frac{Z'e^2}{\left|\vec{r}_i - \vec{l}_j\right|}$ , it can be concluded that:

$$\begin{aligned}
\Delta E_v(x) &= \gamma \int a_n^*\left(\vec{r} - \vec{l}\right) \sum_{i,j,l_j \neq 0} \frac{1}{4\pi\varepsilon_0} \frac{(Z-\mu_v)e^2}{\left|\vec{r}_i - \vec{l}_j\right|} a_n\left(\vec{r} - \vec{l}\right) \mathrm{d}r \sum_l \hat{n}_l \\
&= \gamma \left\langle \Psi_v\left(\vec{r}\right) \right| \sum_{i,j,\vec{R}_j \neq 0} \frac{1}{4\pi\varepsilon_0} \frac{(Z-\mu_v)e^2}{\left|\vec{r}_i - \vec{l}_j\right|} \left| \Psi_v\left(\vec{r}\right) \right\rangle
\end{aligned} \tag{4}$$

**Eq. 4** was obtained by calculating the external field-induced BE ratio, $\gamma$ ,using the known reference values of $\Delta E_v'(x) = E_v(x) - E_v(B)$, $\Delta E_v(B) = E_v(B) - E_v(0)$, and $\Delta E_v(x) = E_v(x) - E_v(0)$ derived from XPS analysis. We obtain.For metallic elements, we have:

$$\gamma = \frac{E_v(x) - E_v(0)}{E_v\left(B\right) - E_v\left(0\right)} \approx \frac{Z_x}{Z_b}\frac{d_b}{d_x} = \left(\frac{Z_b - \mu_v}{Z_b}\right)\left(\frac{d_x}{d_b}\right)^{-1} = \left(\frac{d_x}{d_b}\right)^{-m'} = \frac{E_x}{E_b}, \tag{5}$$

$Z_b$ effective positive charge of bulk atoms, $\mu_v$ is the relative charge shielding factor. $m$ is an indicator for the bond nature of a specific material. The $m'$ is related to the charge shielding factor $\mu_v$ , $m' = 1 - \frac{In \frac{Z_b - \mu_v}{Z_b}}{In\left(\frac{d_x}{d_b}\right)}$ . $\mu_v$ is a very small value, therefore, $m' \approx m = 1$.

For compounds with covalent bonding properties, we use the expression BE ratio, and **Eq.5** can be written as:

$$\begin{cases}
\gamma' = \dfrac{E_v(i) - E_v(0)}{E_v\left(B\right) - E_v\left(0\right)} \\
\gamma' = \gamma^m = \left(\dfrac{E_v(x) - E_v(0)}{E_v\left(B\right) - E_v\left(0\right)}\right)^m \approx \left(\dfrac{Z_x}{Z_b}\dfrac{d_b}{d_x}\right)^m = \left(\dfrac{Z_b - \lambda_v'}{Z_b}\right)^m \left(\dfrac{d_x}{d_b}\right)^{-m} = \left(\dfrac{d_x}{d_b}\right)^{-m'} \approx \left(\dfrac{d_x}{d_b}\right)^{-m} = \dfrac{E_x}{E_b}
\end{cases}$$

.

(6)

$$m'=m\left(1-\frac{In\dfrac{Z_b-\mu_v}{Z_b}}{In\left(\dfrac{d_x}{d_b}\right)}\right)$$, $\mu_v$ is a very small value, therefore, $\dfrac{Z_b-\mu_v}{Z_b}\approx 1$, therefore, $m\approx m'$.

For bond-charge (BC) model，the Hamiltonian of the system becomes:

$$H=\sum_{k\sigma}\frac{\hbar^2k^2}{2m}C_{k\sigma}^{\dagger}C_{k\sigma}+\frac{e_1^2}{2V}\sum_{q}{}^{*}\sum_{\vec{k}\sigma}\sum_{\vec{k}'\lambda}\frac{4\pi}{q^2}C_{\vec{k}+\vec{q},\sigma}^{\dagger}C_{\vec{k}'-\vec{q},\lambda}^{\dagger}C_{\vec{k}'\lambda}C_{\vec{k}\sigma}$$

(7)

$e$ is the basic charge, $e_1=e\Big/\sqrt{4\pi\varepsilon_0}$ . Electronic interactions including charge shielding represented by electron density:

$$\begin{aligned}\hat{V}'_{ee}&=\frac{1}{4\pi\varepsilon_0}\times\frac{e^2}{2\left|\vec{r}-\vec{r}'\right|}\int d^3r\int d^3r'\rho\left(\vec{r}\right)\rho\left(\vec{r}'\right)e^{-\mu(\vec{r}-\vec{r}')}\\&=\frac{1}{2}\int d^3r\int d^3r'V_{ee}(\vec{r}-\vec{r}')C_{\sigma}^{?}(\vec{r})C_{\lambda}^{\dagger}(\vec{r}')C_{\sigma}(\vec{r})C_{\lambda}(\vec{r}')e^{-\mu(\vec{r}-\vec{r}')}\end{aligned}$$

(8)

and

$$\begin{cases}\rho\left(\vec{r}\right)=C_{\sigma}^{+}\left(\vec{r}\right)C_{\sigma}\left(\vec{r}\right)\\C_{\sigma}^{\dagger}(\vec{r})=\sum_{\vec{R}}\psi_{\vec{R}}^{*}(\vec{r})C_{\vec{R}\sigma}^{?}=\sum_{i}\psi_{\vec{R}i}^{*}(\vec{r})C_{i\sigma}^{?}\\C_{\vec{k}\sigma}^{\dagger}=\frac{1}{\sqrt{N}}\sum_{i}e^{i\vec{k}\vec{r}_i}C_{i\sigma}^{\dagger}\end{cases}$$

(9)

Then, we have

$$
\begin{cases}
\hat{V}'_{ee} = \dfrac{e_1^2}{2} \sum_{l',\sigma'} \sum_{m',\lambda'} \sum_{l,\sigma} \sum_{m,\lambda} (\vec{k}_{l'}\sigma', \vec{k}_{m'}\lambda' \left| \dfrac{e^{-\mu(\vec{r}-\vec{r}')}}{|r-r'|} \right| \vec{k}_l\sigma, \vec{k}_m\lambda) C^{\dagger}_{\vec{k}_{l'}\sigma'} C^{\dagger}_{\vec{k}_{m'}\lambda'} C_{\vec{k}_m\lambda} C_{\vec{k}_l\sigma} \\
\quad = \dfrac{e_1^2}{2} \sum_{l'} \sum_{m'} \sum_{l,\sigma} \sum_{m,\lambda} \delta_{\sigma'\sigma} \delta_{\lambda'\lambda} C^{\dagger}_{\vec{k}_{l'}} C^{\dagger}_{\vec{k}_{m'}} (\vec{k}_{l'}, \vec{k}_{m'} \left| \dfrac{e^{-\mu(\vec{r}-\vec{r}')}}{|r-r'|} \right| \vec{k}_l, \vec{k}_m) C_{\vec{k}_m} C_{\vec{k}_l} \\
\quad = \dfrac{e_1^2}{2V} \sum_{\vec{k}} \sum_{\vec{k}'} \sum_{\vec{q}} \sum_{\sigma\lambda} \dfrac{4\pi}{q^2+\mu^2} C^{\dagger}_{\vec{k}+\vec{q},\sigma} C^{\dagger}_{\vec{k}'-\vec{q},\lambda} C_{\vec{k}'\lambda} C_{\vec{k}\sigma} \\
(\vec{k}_{l'}, \vec{k}_{m'} \left| \dfrac{e^{-\mu(\vec{r}-\vec{r}')}}{|r-r'|} \right| \vec{k}_l, \vec{k}_m) = \delta(\vec{k}_{l'} + \vec{k}_{m'}, \vec{k}_l + \vec{k}_m) \dfrac{1}{V} \dfrac{4\pi}{\left(\vec{k}_{l'} - \vec{k}_l\right) + \mu^2} (\vec{k}_l = \vec{k}, \vec{k}_m = \vec{k}', \vec{k}_{l'} = \vec{k} + \vec{q}, \vec{k}_{m'} = \vec{k}' - \vec{q})
\end{cases}
$$

(10)

The electron interaction term without charge shielding for density fluctuations ($\mu=0$):

$$
\hat{V}_{ee} = \frac{1}{4\pi\varepsilon_0} \frac{e^2}{2|\vec{r}-\vec{r}'|} \int d^3r \int d^3r' \rho(\vec{r}) \rho(\vec{r}') = \frac{e_1^2}{2V} \sum_q {}^{*} \sum_{\vec{k}\sigma} \sum_{\vec{k}'\lambda} \frac{4\pi}{q^2} C^{?}_{\vec{k}+\vec{q},\sigma} C^{\dagger}_{\vec{k}'-\vec{q},\lambda} C_{\vec{k}'\lambda} C_{\vec{k}\sigma}
$$

(11)

The formula for electron-electron interaction potential is:

$$
\begin{aligned}
\delta V_{bc} &= \hat{V}'_{ee} - \hat{V}_{ee} \\
&= \frac{e_1^2}{2V} \sum_{\vec{k}} \sum_{\vec{k}'} \sum_{\vec{q}} \sum_{\sigma\lambda} \frac{4\pi}{q^2+\mu^2} C^{\dagger}_{\vec{k}+\vec{q},\sigma} C^{\dagger}_{\vec{k}'-\vec{q},\lambda} C_{\vec{k}'\lambda} C_{\vec{k}\sigma} - \frac{e_1^2}{2V} \sum_q {}^{*} \sum_{\vec{k}\sigma} \sum_{\vec{k}'\lambda} \frac{4\pi}{q^2} C^{\dagger}_{\vec{k}+\vec{q},\sigma} C^{\dagger}_{\vec{k}'-\vec{q},\lambda} C_{\vec{k}'\lambda} C_{\vec{k}\sigma} \\
&= \frac{1}{4\pi\varepsilon_0} \frac{e^2}{2|\vec{r}-\vec{r}'|} \int d^3r \int d^3r' \rho(\vec{r}) \rho(\vec{r}') e^{-\mu(\vec{r}-\vec{r}')} - \frac{1}{4\pi\varepsilon_0} \frac{e^2}{2|\vec{r}-\vec{r}'|} \int d^3r \int d^3r' \rho(\vec{r}) \rho(\vec{r}')
\end{aligned}
$$

(12)

Deformation bond energy $\delta V_{bc}$ can be represented as:

$$
\delta V_{bc} = \frac{1}{4\pi\varepsilon_0} \frac{e^2}{2|\vec{r}-\vec{r}'|} \int d^3r \int d^3r' \delta\rho(\vec{r}) \delta\rho(\vec{r}')
$$

(13)

The formation of chemical bonds is related to fluctuations in electron density $\delta\rho$. The charge density is in the same units as those used by CASTEP during calculations. The values are normalized to the number of electrons per cell. The density fields displayed in the 3D model are, however, normalized to the number of electrons per Å$^3$.

Finally, we have the Hartree–Fock ground state energy formula:

$$\frac{E_0}{N} = \frac{3}{5}\frac{\hbar^2 k_F^2}{2m} - \frac{3e^2}{4\pi}k_F + \frac{e^2}{2a_H}\varepsilon_c\left(r_s\right)$$
$$= \frac{e^2}{2a_H}\left(\frac{2.21}{r_s^2} + \varepsilon_x\left(r_s\right) + \varepsilon_c\left(r_s\right)\right) \tag{14}$$

with the exchange correlation energy $\varepsilon_{xc}\left(r_s\right)$ [18, 19]:

$$\begin{cases} \varepsilon_{xc}\left(r_s\right) = \varepsilon_x\left(r_s\right) + \varepsilon_c\left(r_s\right) \\ \varepsilon_c\left(r_s\right) = \begin{cases} -0.2846 / \left(1 + 1.0529\sqrt{r_s} + 0.3334 r_s\right)\left(r_s \geq 1\right) \\ -0.0960 + 0.0622 \ln r_s - 0.0232 r_s + 0.0040 r_s \ln r_s \left(r_s \leq 1\right) \end{cases} \\ \varepsilon_x\left(r_s\right) = -0.9164 / r_s \end{cases} \tag{15}$$

$\varepsilon_c\left(r_s\right)$ represents the correlation energy and $\varepsilon_x\left(r_s\right)$ the exchange energy.

## Results and discussion

### Electronic structure

The geometric structures of $MoS_2$, $MoSe_2$, and $MoTe_2$ are illustrated in **Fig. 1**. The optimized lattice parameters are listed in Table 1. For MoS2, the lattice parameters in a, b, and c are 6.385, 6.385, and 6.385 Å, respectively, and in α, β, and γ angles are α = 90.00°, β = 90.00, and γ = 120.00°. In Fig. 2, the DOS of MoS2, MoSe2, and MoTe2 are given. In the conductive band, the main peaks of the electron density distribution occur at −2.06 eV, −1.83 eV, and −1.46 eV for $MoS_2$, $MoSe_2$, and $MoTe_2$, respectively. The partial DOS of s, p, and d orbitals were analyzed for $MoS_2$, $MoSe_2$, and $MoTe_2$. The d orbitals contribute the vast majority of the electrons in the valence and conduction bands.

We calculated the band structures of $MoS_2$, $MoSe_2$, and $MoTe_2$ as shown in **Fig. 2**. The band gap values of $MoS_2$, $MoSe_2$, and $MoTe_2$ structures are 1.374 eV, 1.426 eV, and 1.016 eV, respectively. Structurally, $MoS_2$, $MoSe_2$, and $MoTe_2$ are all indirect bandgap semiconductor materials with adjustable optical bandgaps in the 1.0–1.5 eV range that allow them to absorb visible and near-infrared light. **Fig. 3** exhibits the deformation charge density obtained by analyzing the calculated results of the $MoS_2$, $MoSe_2$, and $MoTe_2$ structures, where the electron distributions are indicated by the color scale. The blue and red areas represent increases and decreases of the electron distribution, respectively; the electrons are convergent in the positive region and divergent in the negative region. Thus, the blue regions indicate a gathering of electrons, yielding a positive deformation charge density, while the red regions indicate a divergence of electrons and a resultant negative deformation charge density.

The established BBC model was used to convert the values of the Hamiltonian into the bonding values (Table 3). Using **Eq. 13**, the deformation bond energies are −0.4851 eV, −1.2864 eV, and −0.7692 eV for $MoS_2$, $MoSe_2$, and $MoTe_2$, respectively.

### Bond properties

The lattice density formula for a three-dimensional lattice is (see support information) [20]:

$$\rho_0(E)=\frac{2\theta(b-|E-\varepsilon_0|)}{\pi b^2}\sqrt{b^2-(E-\varepsilon_0)^2} \tag{16}$$

where $b$ is the full width at half maximum (FWHM). When $|E-\varepsilon_0|<b$, we have $\theta(b-|E-\varepsilon_0|)=1$.

Considering the core energy level $E=E_v(x)$, we obtain:

$$\rho_0(E)=\frac{2}{\pi b^2}\sqrt{b^2-(E_v(x)-E_v(I))^2} \tag{17}$$

Through XPS measurements, we can obtain the BE of the electrons and calculate the BE ratio. The FWHM, BE, and BE ratio of Mo, $MoS_2$, $MoSe_2$, and $MoTe_2$ are shown in **Table 4**. **Fig. 4** shows the lattice density of the BE of electrons at different core levels calculated by **Eq. 17**. Therefore, as long as we know the BE and FWHM, we can obtain the lattice density distribution of the core level.

The BE ratio is used as the input for the calculation of bond properties. By combining BOLS theory[21] with the BBC model, we further obtain[22]:

$$\begin{cases} E_v(I)=E_v(0)+\Delta E_v(B)\gamma^m & \text{(binding energy shift)} \\ \varepsilon_x=\gamma^{-1}-1 & \text{(local bond strain)} \\ \delta E_C=z_x E_x/z_b E_b-1=z_{xb}\gamma^m-1 & \text{(atomic cohesive energy)} \\ \delta E_D=(E_x/d_x^3)/(E_b/d_b^3)-1=\gamma^{m+3}-1 & \text{(bond energy density)} \\ z_x=\dfrac{12}{\{8\ln(2\gamma-1)+1\}} & \text{(atomic coordination number)} \\ t=b/z_x & \text{(overlapping integral)} \end{cases} \tag{18}$$

For the bulk Mo atom, the energy level shift is $\Delta E_v(B)$ = 2.707 eV and the single atom energy level shift is $E_v(0)$ = 224.862 eV[23]. For single-metal materials, $m$ = 1, while for compounds, there will be different values of the bond nature indicator $m$. For the same compound, the value of m is the same.Using **Eq. 18**, we calculate the local bond strain(LBS) $\varepsilon_x$ (%), relative atomic cohesive energy(ACE) $\delta E_C$ (%), relative bond energy density(BED) $\delta E_D$ (%), BE ratio $\gamma$, coordination number (CN) $z_x$, and FWHM $b$ of Mo, $MoS_2$, $MoSe_2$, and $MoTe_2$. For companies using the BBC model, the XPS energy shift can be used to calculate the performance of chemical bonds.

From **Fig. 5**, we see that the BE, ACE, and BED are related to the BE ratio. If the BE ratio $\gamma$ is constant, we find that the higher the value of $m$, the greater the BE and BED. The ACE of $MoS_2$, $MoSe_2$, and $MoTe_2$ is larger than that of elemental Mo. Our calculations provide a theoretical reference for the bonding properties of $MoS_2$, $MoSe_2$, and $MoTe_2$ semiconductor materials. We use these results to develop a method for the accurate analysis of the bonding properties of semiconductor materials.

### Electrostatic shielding

Considering the energy scaling *L* for the potential function $V\left(\vec{r}\right)$, we have

$$L=\left|\frac{V_i\left(\vec{r_{ij}}\right)}{V_0\left(\vec{r_{ij}}\right)}\right|=\mathrm{e}^{-\mu_i r_{ij}} .$$

(19)

The electrostatic shielding parameter $\mu$ in **Eq. 19** is related to the electrostatic shielding effect.

According to **Eq. 19**, further, we can obtain the bonding potential energy $\left(\delta V_{bc}<0\right)$ and exchange correlation energy from the following relationship:

$$\left(13.6\varepsilon_c\left(r_s\right)+\delta V_{bc}\right)=13.6\varepsilon_c\left(r_s\right)e^{-\mu_{A1}r_{ij}}$$

(20)

$$\left(13.6\varepsilon_c\left(r_s\right)+\delta V_{bc}\right)=\delta V_{bc}e^{-\mu_{A2}r_{ij}}$$

(21)

$$t=-\delta V_{bc}e^{-\mu_{A3}r_{ij}}$$

(22)

The antibonding potential energy $\left(\delta V_{bc}>0\right)$ and exchange correlation energy have the following relationship:

$$\left|\left(13.6\varepsilon_c\left(r_s\right)+\delta V_{bc}\right)\right|=\left|13.6\varepsilon_c\left(r_s\right)\right|e^{-\mu_{B1}r_{ij}}$$

(23)

$$\left|\left(13.6\varepsilon_c\left(r_s\right)+\delta V_{bc}\right)\right|=\delta V_{bc}e^{-\mu_{B2}r_{ij}}$$

(24)

$$t=\delta V_{bc}e^{-\mu_{B3}r_{ij}}$$

(25)

**Fig. 6** shows the influence of electrostatic shielding effect on the deformation bond energy calculated by **Eq. 19** and **Eq. 21**. **Eq. 20** represents the effect of deformation bond energy $\delta V_{bc}$ on the ground state energy under different values of electrostatic shielding $\mu_{A1}$. **Eq. 21** represents the effect of deformation bond energy $\delta V_{bc}$ on density fluctuations under electrostatic shielding $\mu_{A2}$; $t=b/z_x$ is the overlapping integral. **Eq. 22** allows *us to calculate the electrostatic shielding $\mu_{A3}$ values for $MoS_2$, $MoSe_2$, and $MoTe_2$, as shown in **Table 4**.*

*The potential energy surface and phase operator can be calculated as follows[24]:*

$$\begin{cases} V_i\left(\overrightarrow{r_{ij}}\right)=V_0\left(\overrightarrow{r_{ij}}\right)\mathrm{e}^{-\mu_i r_{ij}}\cos k_1 a \\ V_i\left(\overrightarrow{r_{ij}}\right)=V_0\left(\overrightarrow{r_{ij}}\right)\mathrm{e}^{-\mu_i r_{ij}}(\cos k_1 a+\cos k_2 a) \\ V_i\left(\overrightarrow{r_{ij}}\right)=V_0\left(\overrightarrow{r_{ij}}\right)\mathrm{e}^{-\mu_i r_{ij}}(\cos k_1 a+\cos k_2 a+\cos k_3 a) \end{cases}$$

*(26)*

*Here, $k_1$, $k_2$, and $k_3$ represent the components of **k** in the x, y, and z directions, respectively. a is the lattice constant.*

This yields the tight-binding approximations for one-dimensional (1D) lattices:

$$\begin{cases} E(k)=\varepsilon_0+2t\cos k_1 a=\varepsilon_0+2\delta V_{bc}e^{-\mu_i r_{ij}}\cos k_1 a \\ E_{band}=2t\cos k_1 a=2\delta V_{bc}e^{-\mu_i r_{ij}}\cos k_1 a \end{cases}$$

(27)

2D lattices:

$$\begin{cases} E(k)=\varepsilon_0+2t\left(\cos k_1 a+\cos k_2 a\right)=\varepsilon_0+2\delta V_{bc}e^{-\mu_i r_{ij}}\left(\cos k_1 a+\cos k_2 a\right) \\ E_{band}=2t\left(\cos k_1 a+\cos k_2 a\right)=2\delta V_{bc}e^{-\mu_i r_{ij}}\left(\cos k_1 a+\cos k_2 a\right) \end{cases}$$

(28)

and 3D lattices:

$$\begin{cases} E(k)=\varepsilon_0+2t\left(\cos k_1 a+\cos k_2 a+\cos k_3 a\right)=\varepsilon_0+2\delta V_{bc}e^{-\mu_i r_{ij}}\left(\cos k_1 a+\cos k_2 a+\cos k_3 a\right) \\ E_{band}=2t\left(\cos k_1 a+\cos k_2 a+\cos k_3 a\right)=2\delta V_{bc}e^{-\mu_i r_{ij}}\left(\cos k_1 a+\cos k_2 a+\cos k_3 a\right) \end{cases}$$

(29)

Using **Eq. 28** and **Eq. 29** and the overlaying integral *t=b/z*, we can calculate the 2D and 3D energy bands of $MoS_2$, $MoSe_2$, and $MoTe_2$, as shown in **Fig. 7**.The fluctuations in charge density can be represented by two-dimensional and three-dimensional band diagrams.

## Conclusion

We computationally analyzed the electronic structure and bond properties of $MoS_2$, $MoSe_2$, and $MoTe_2$ materials and found that electrostatic shielding by electron exchange is the main cause of density fluctuations. The fluctuations in charge density can be represented by two-dimensional and three-dimensional band diagrams. The density of the Green's function can be calculated from the energy shifts. the XPS energy shift can be used to calculate the performance of chemical bonds. These new methods and ideas can be applied in photoelectron spectroscopy to the study of the atomic bonding, BE shifts, and electronic structure of materials.

**Keywords: Electrostatic shielding effect • Binding energy shift • X-ray photoelectron spectroscopy • Density functional theory**

## References

[1] H. Li, Q. Zhang, C. C. R. Yap, B. K. Tay, T. H. T. Edwin, A. Olivier, D. Baillargeat *Advanced Functional Materials*. **2012**, *22*, 1385-1390.

[2] Z. Wang, B. Mi *Environmental Science & Technology*. **2017**, *51*, 8229-8244.

[3] H. Li, J. Wu, Z. Yin, H. Zhang *Accounts of Chemical Research*. **2014**, *47*, 1067-1075.

[4] A. Splendiani, L. Sun, Y. Zhang, T. Li, J. Kim, C.-Y. Chim, G. Galli, F. Wang *Nano Letters*. **2010**, *10*, 1271-1275.

[5] R. Ganatra, Q. Zhang *ACS Nano*. **2014**, *8*, 4074-4099.

[6] D. Kong, H. Wang, J. J. Cha, M. Pasta, K. J. Koski, J. Yao, Y. Cui *Nano Letters*. **2013**, *13*, 1341-1347.

[7] X. Wang, Y. Gong, G. Shi, W. L. Chow, K. Keyshar, G. Ye, R. Vajtai, J. Lou, Z. Liu, E. Ringe, B. K. Tay, P. M. Ajayan *ACS Nano*. **2014**, *8*, 5125-5131.

[8] A. Eftekhari *Applied Materials Today*. **2017**, *8*, 1-17.

[9] C. Ruppert, B. Aslan, T. F. Heinz *Nano Letters*. **2014**, *14*, 6231-6236.

[10] S. Cho, S. Kim, J. H. Kim, J. Zhao, J. Seok, D. H. Keum, J. Baik, D.-H. Choe, K. J. Chang, K. Suenaga, S. W. Kim, Y. H. Lee, H. Yang *Science*. **2015**, *349*, 625-628.

[11] N. R. Pradhan, D. Rhodes, S. Feng, Y. Xin, S. Memaran, B.-H. Moon, H. Terrones, M. Terrones, L. Balicas *ACS Nano*. **2014**, *8*, 5911-5920.

[12] D. Voiry, M. Salehi, R. Silva, T. Fujita, M. Chen, T. Asefa, V. B. Shenoy, G. Eda, M. Chhowalla *Nano Letters*. **2013**, *13*, 6222-6227.

[13] S. Larentis, B. Fallahazad, E. Tutuc *Applied Physics Letters*. **2012**, *101*.

[14] X. Luo, F. C. Chen, J. L. Zhang, Q. L. Pei, G. T. Lin, W. J. Lu, Y. Y. Han, C. Y. Xi, W. H. Song, Y. P. Sun *Applied Physics Letters*. **2016**, *109*.

[15] K. Deng, G. Wan, P. Deng, K. Zhang, S. Ding, E. Wang, M. Yan, H. Huang, H. Zhang, Z. Xu, J. Denlinger, A. Fedorov, H. Yang, W. Duan, H. Yao, Y. Wu, S. Fan, H. Zhang, X. Chen, S. Zhou *Nature Physics*. **2016**, *12*, 1105-1110.

[16] H. Qiu, H. Li, J. Wang, Y. Zhu, M. Bo *physica status solidi (RRL)–Rapid Research Letters*. **2022**, *16*, 2100444.

[17] J. P. Perdew, K. Burke, M. Ernzerhof *Physical review letters*. **1996**, *77*, 3865.

[18] M. Gell-Mann, K. A. Brueckner *Physical Review*. **1957**, *106*, 364-368.

[19] J. P. Perdew, A. Zunger *Physical Review B*. **1981**, *23*, 5048-5079.

[20] W. A. Schwalm, M. K. Schwalm *Physical Review B*. **1988**, *37*, 9524-9542.

[21] C. Q. Sun *Progress in Solid State Chemistry*. **2007**, *35*, 1-159.

[22] M. Bo, Y. Guo, X. Yang, J. He, Y. Liu, C. Peng, Y. Huang, C. Q. Sun *Chemical Physics Letters*. **2016**, *657*, 177-183.

[23] W. Zhou, M. Bo, Y. Wang, Y. Huang, C. Li, C. Q. Sun *RSC Advances*. **2015**, *5*, 29663-29668.

[24] Z. Wang, Y. Huang, F. Li, Y. Chuang, Z. Huang, M. Bo *Chemical Physics Letters*. **2022**, *808*, 140124.

[25] M. A. Baker, R. Gilmore, C. Lenardi, W. Gissler *Applied Surface Science*. **1999**, *150*, 255-262.

[26] L. Wang, J. Zhao, X. Tang, S. Kuang, L. Qin, H. Lin, Q. Li *Surfaces and Interfaces*. **2023**, *39*, 102956.

[27] K. Ueno, K. Fukushima *Applied Physics Express*. **2015**, *8*, 095201.

[28] X. Yang, Z. Dong, C. Q. Sun *Applied Physics Letters*. **2023**, *123*, 053101.

## Table and Figure captions:

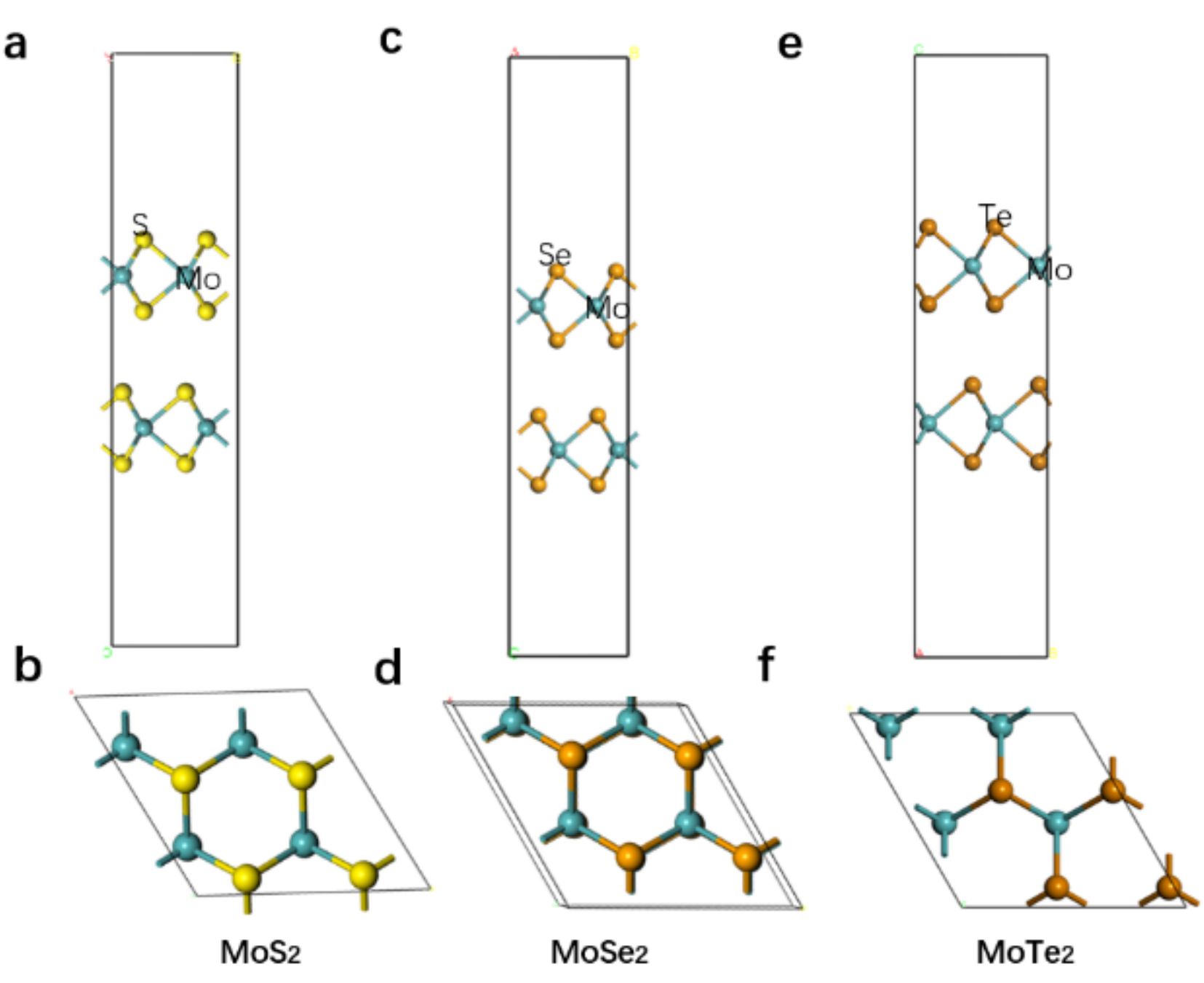

**Fig. 1** Top (a) and side (b) views of $MoS_2$. Top (c) and side (d) views of $MoSe_2$. Top (e) and side (f) views of $MoTe_2$.

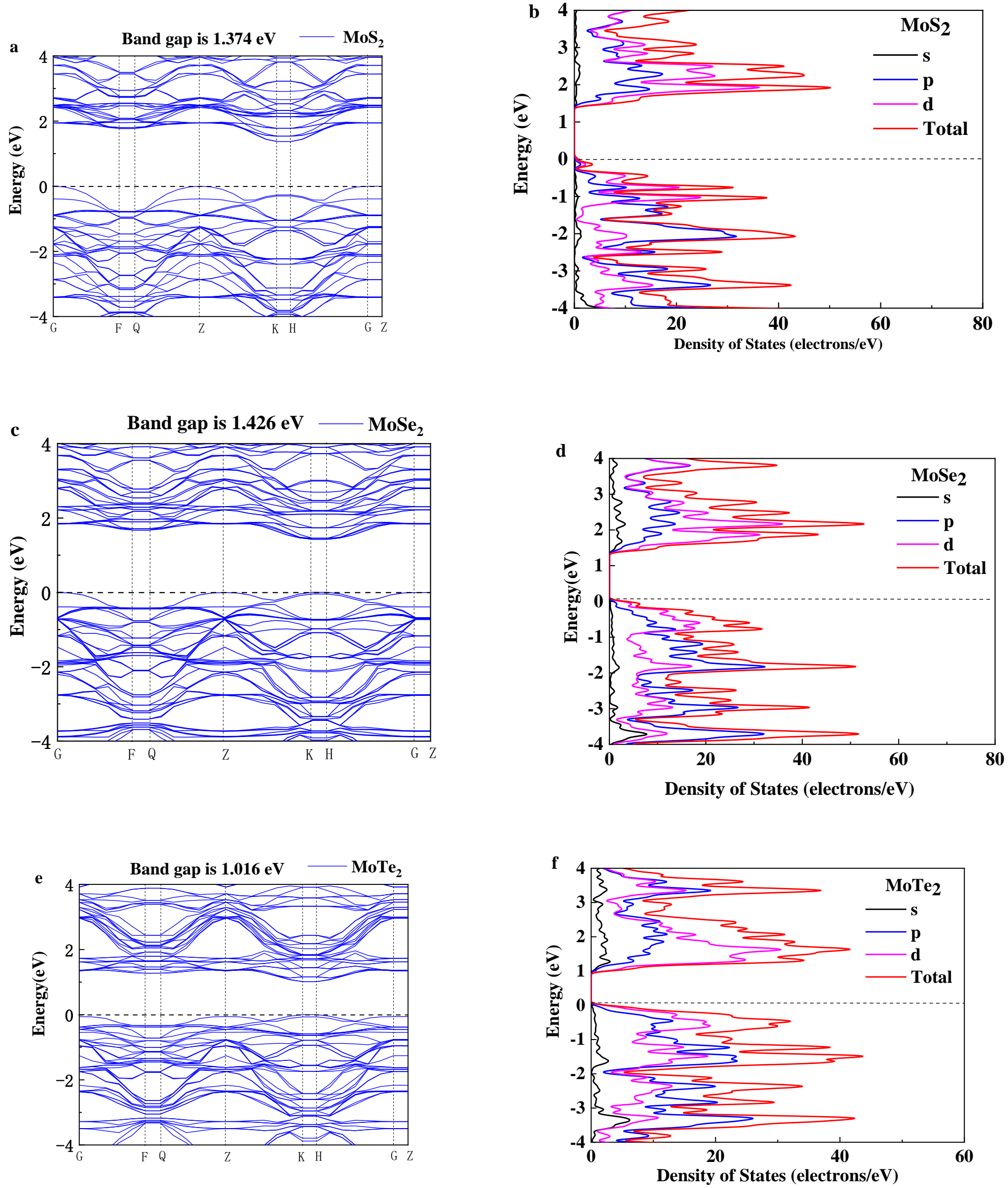

**Fig. 2** Band structure of (a) $MoS_2$, (c) $MoSe_2$, and (e) $MoTe_2$. *s*-, *p*-, and *d*-orbital partial DOS of (b) $MoS_2$, (d) $MoSe_2$, and (f) $MoTe_2$.

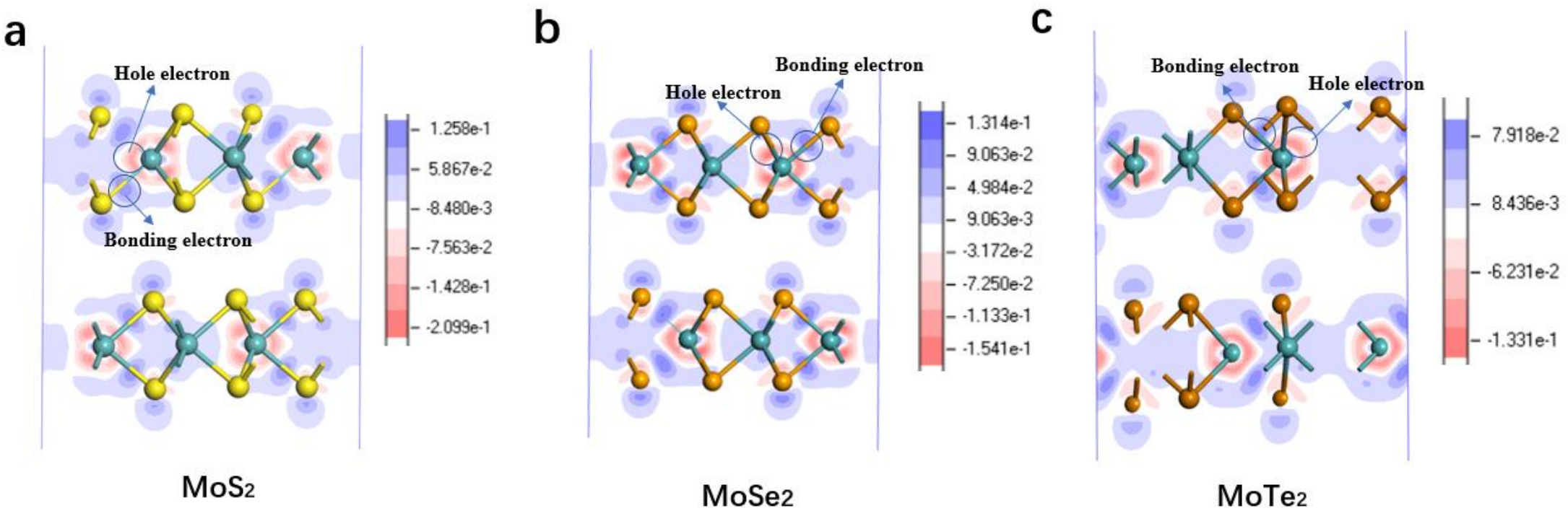

**Fig. 3** Deformation charge densities of (a) $MoS_2$, (b) $MoSe_2$, and (c) $MoTe_2$.

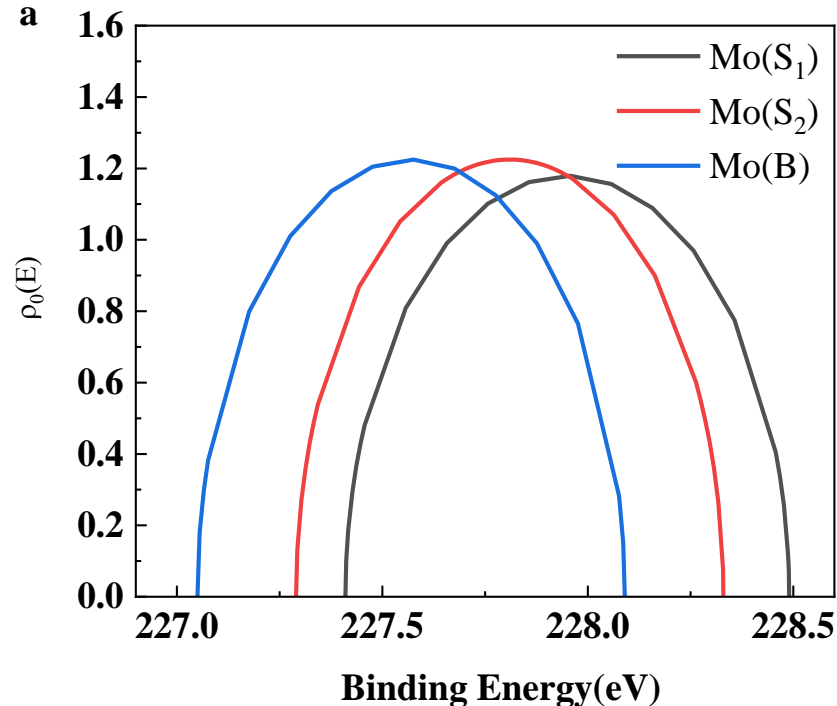

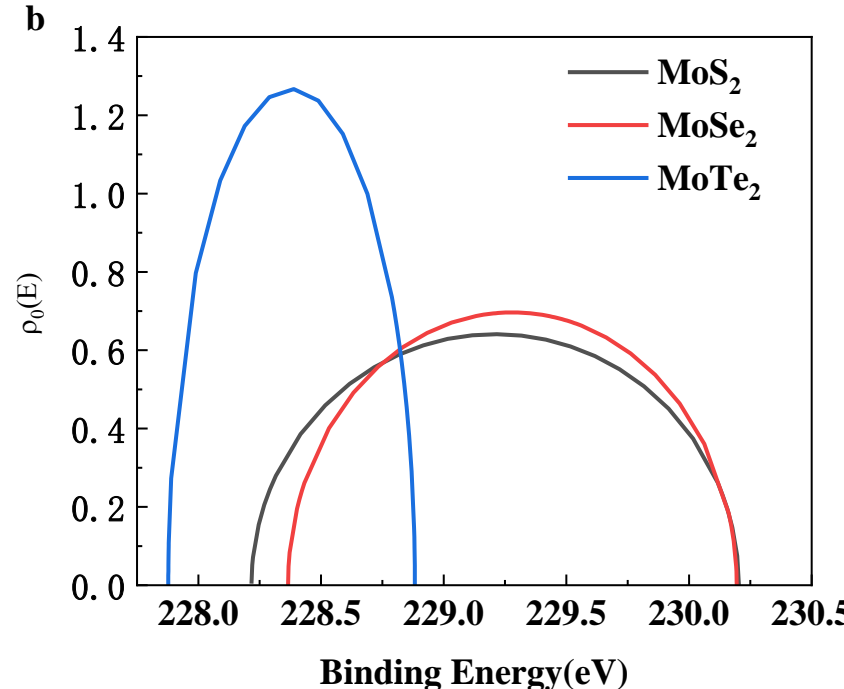

**Fig. 4** Lattice density and BE shift of (a) Mo (100) surface[23] with three components representing the bulk B, $S_2$, and $S_1$ states. (b) $MoS_2$[25], $MoSe_2$[26], and $MoTe_2$[27].

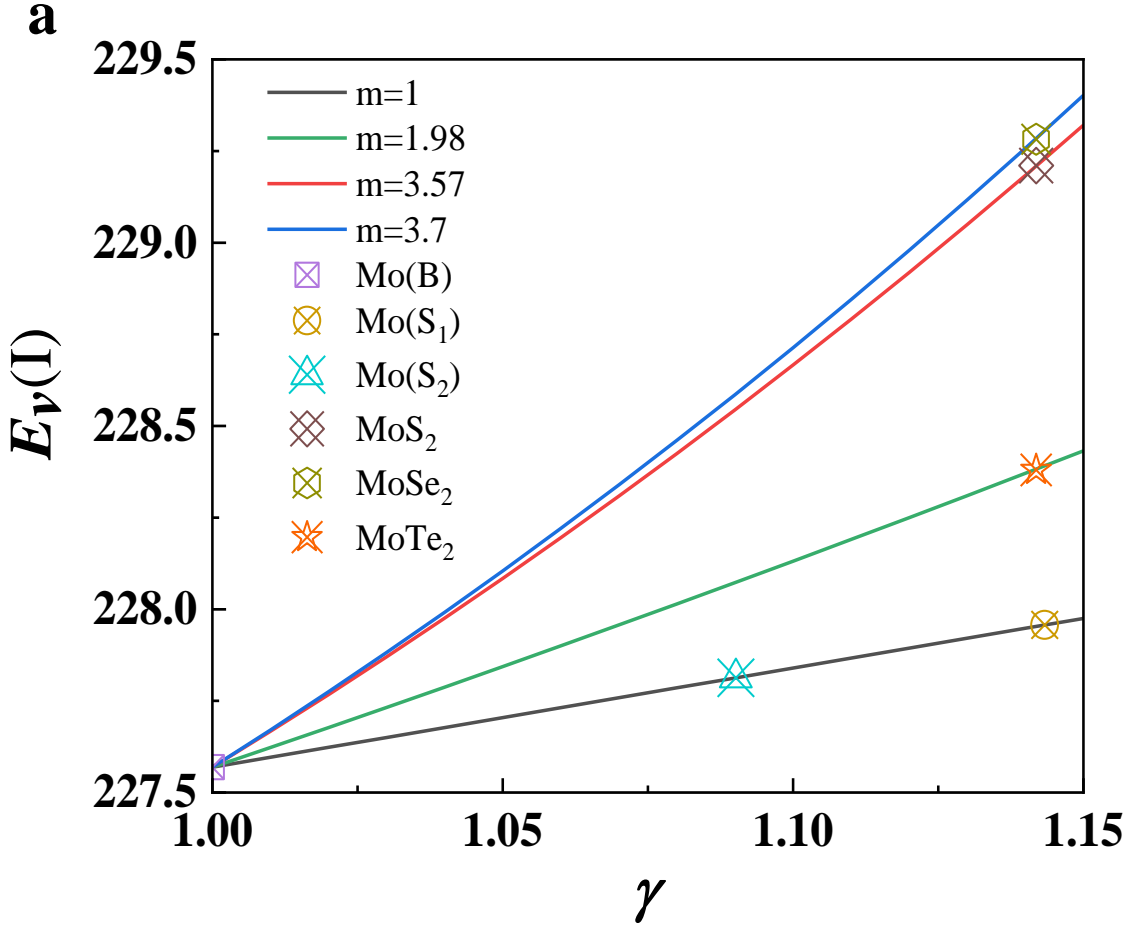

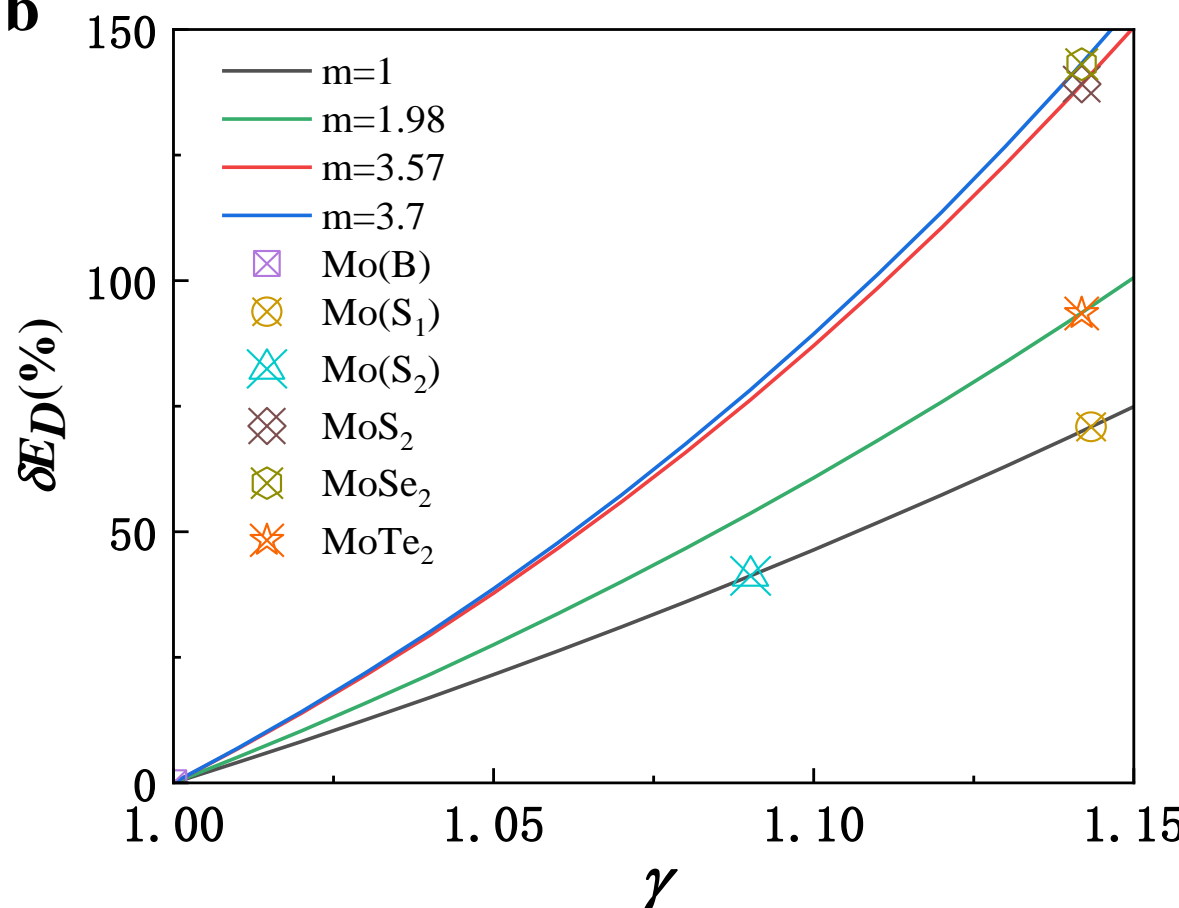

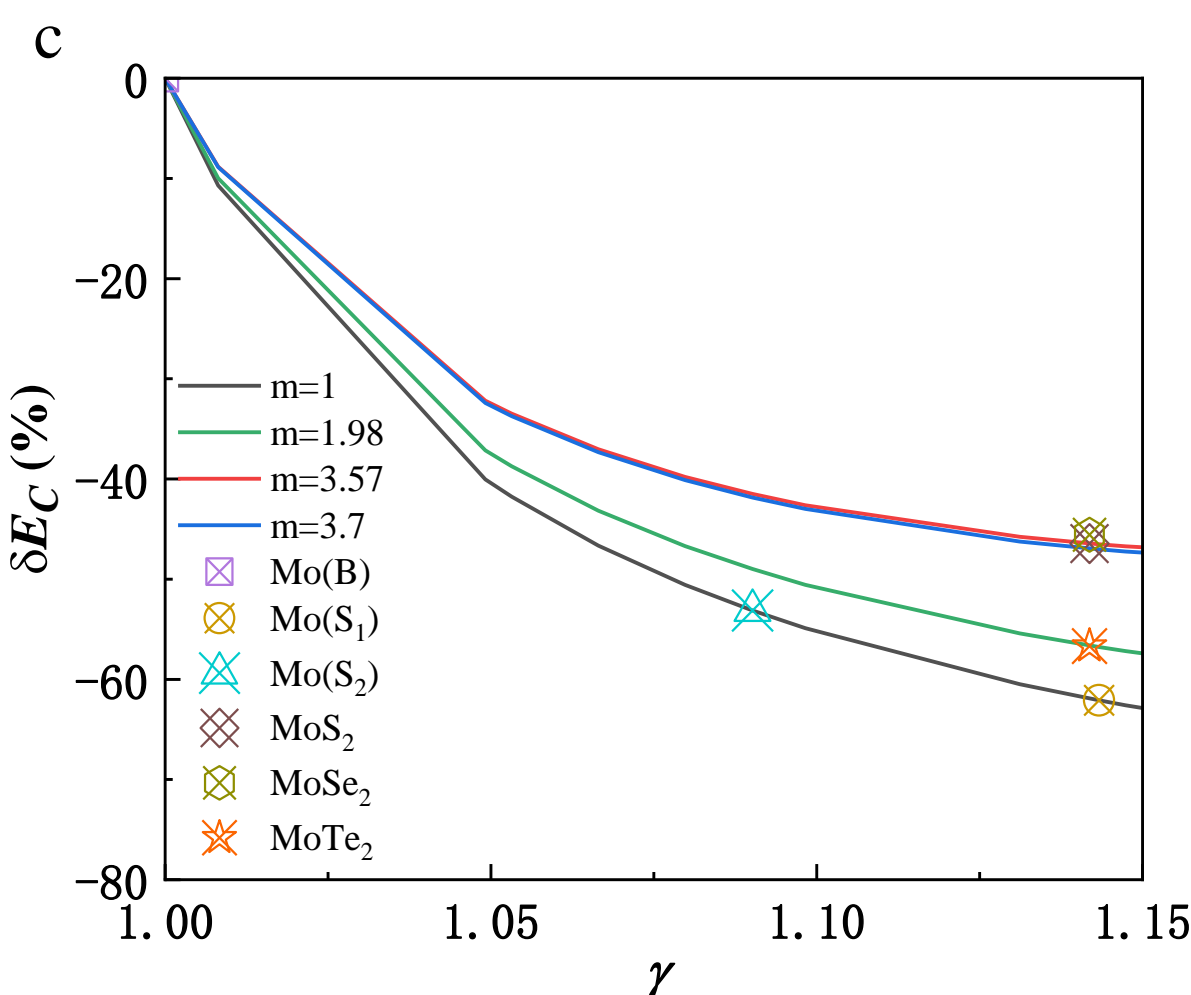

**Fig. 5** BE ratio $\gamma$ versus (a) relative ACE $\delta E_C$ (%), (b) relative BED $\delta E_D$ (%). *m* is an indicator of the bond nature. The related derived information is listed in **Table 4**.

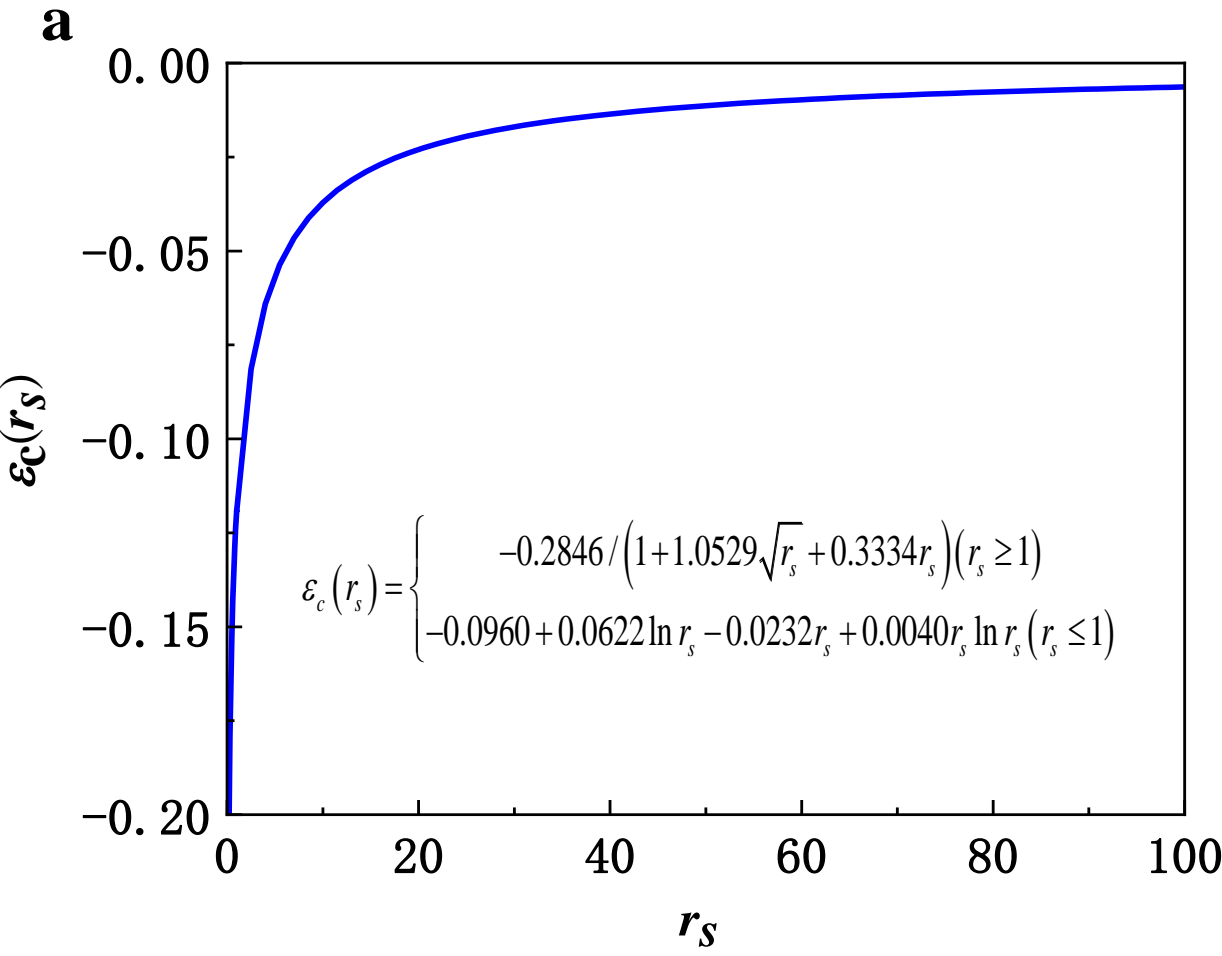

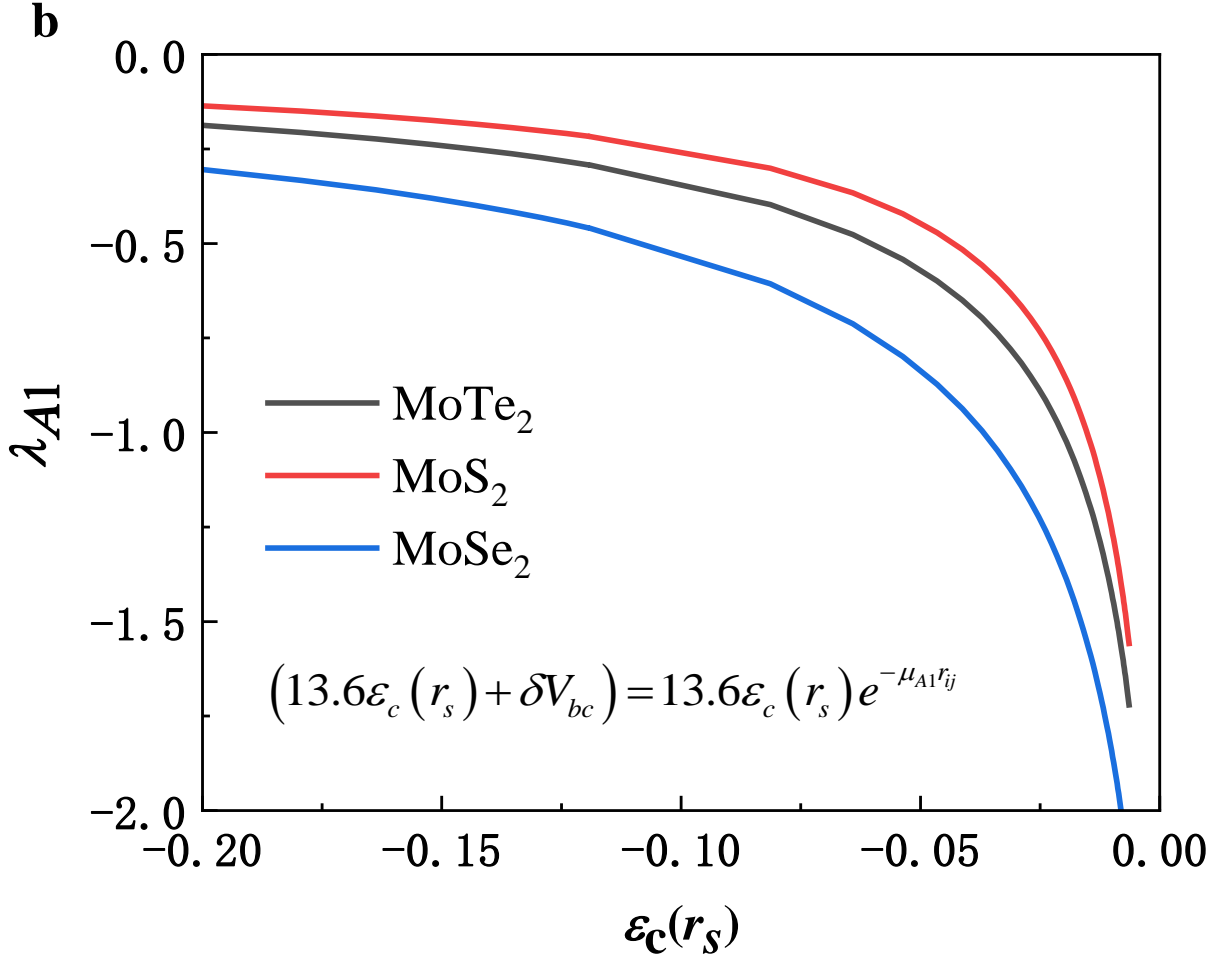

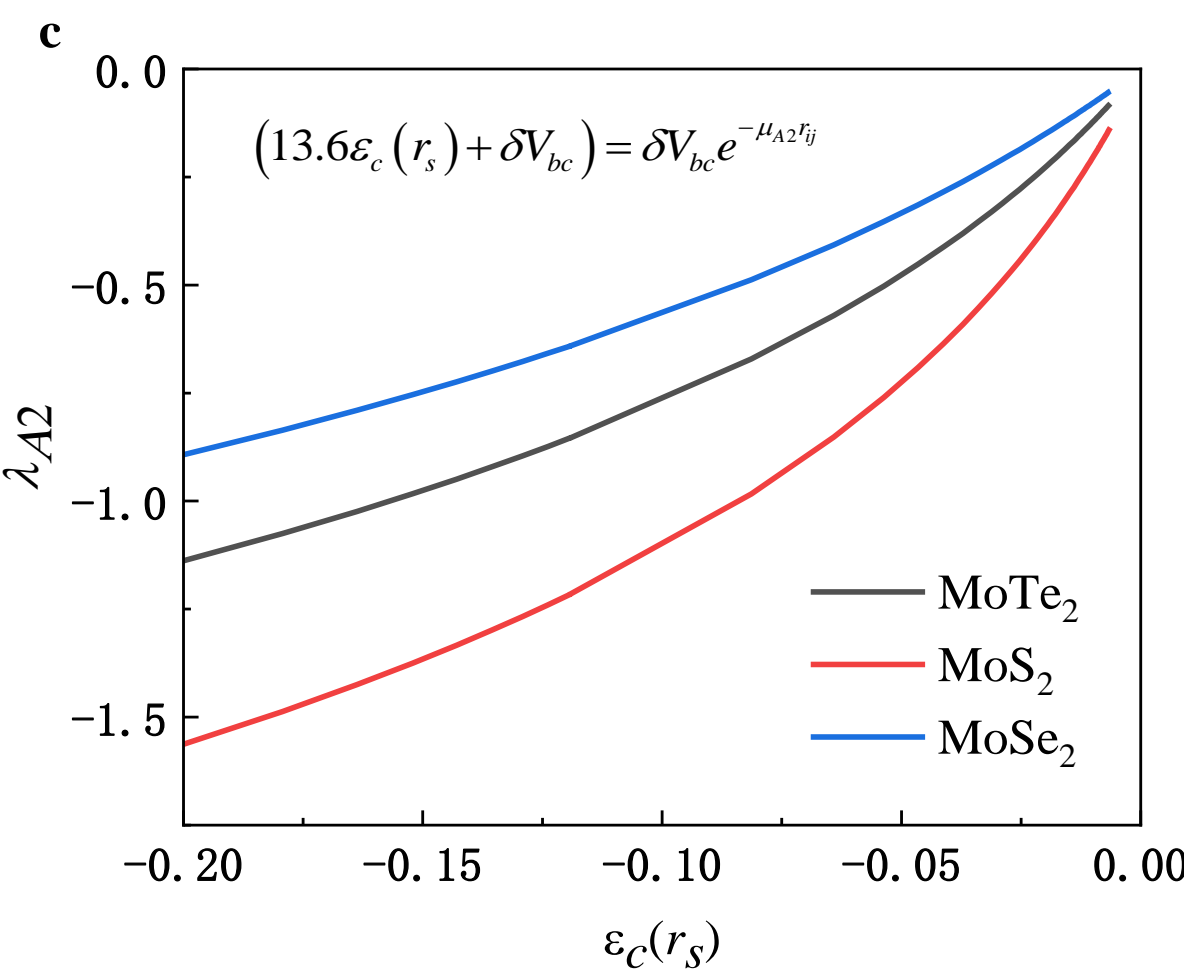

**Fig. 6** Electrostatic shielding effect and energy scaling of $MoS_2$, $MoSe_2$, and $MoTe_2$

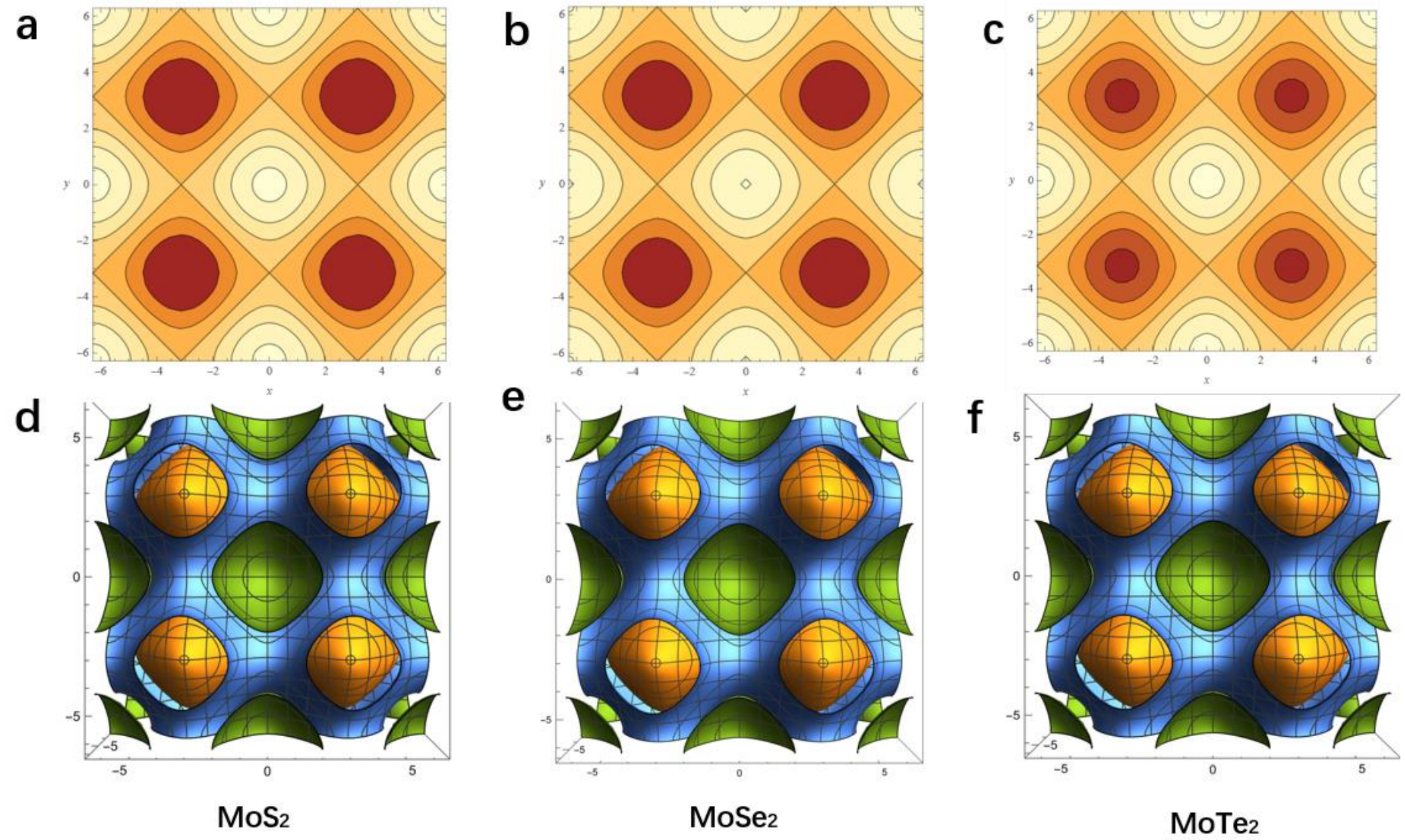

**Fig. 7** 2D (a) $MoS_2$, (b) $MoSe_2$, (c) $MoTe_2$, and 3D (a) $MoS_2$, (b) $MoSe_2$, (c) $MoTe_2$ energy band diagrams obtained with the overlapping integral *t*.

**Table 1** Lattice parameters of $MoS_2$, $MoSe_2$, and $MoTe_2$.

| | Angle | | | Lattice parameters | | |
|---|---|---|---|---|---|---|
| | α | β | γ | *a* (Å) | *b* (Å) | *c* (Å) |
| $MoS_2$ | 90° | 90° | 120° | 6.385 | 6.385 | 26.000 |
| $MoSe_2$ | 90° | 90° | 120° | 7.123 | 7.123 | 28.000 |

| $MoTe_2$ | 90° | 90° | 120° | 6.643 | 6.643 | 29.000 |
|---|---|---|---|---|---|---|

**Table 2** Cut-off energies, band gaps, and *k*-points of $MoS_2$, $MoSe_2$, and $MoTe_2$.

| Element | Cut-off energy | *k*-point | Band gap |
|---|---|---|---|
| $MoS_2$ | 421.8 eV | 8×8×1 | 1.374 eV |
| $MoSe_2$ | 421.8 eV | 8×8×1 | 1.426 eV |
| $MoTe_2$ | 421.8 eV | 8×8×1 | 1.023 eV |

**Table 3** Density of deformation charge $\delta\rho\left(\vec{r_{ij}}\right)$, electrostatic shielding parameter $\lambda_{A3}$, and the deformation bond energy $\delta V_{bc}\left(\vec{r_{ij}}\right)$, as obtained from the BBC model.

$\left(\varepsilon_0 = 8.85\times10^{-12}C^2N^{-1}m^{-2}, e = 1.60\times10^{-19}C, \left|\vec{r_{ij}}\right| \approx d_{ij}/2\right)$

| | $MoS_2$ | $MoSe_2$ | $MoTe_2$ |
|---|---|---|---|
| $r_{ij}$ ( Å ) | 1.208 | 1.272 | 1.328 |
| $r_i$ ( Å ) | 1.54（Mo） | 1.54（Mo） | 1.54（Mo） |
| $\vec{r}_j$ ( Å ) | 1.05(S) | 1.2(Se) | 1.38(Te) |
| $\delta\rho^{Hole-electron}\left(\vec{r_{ij}}\right)\left(e/Å^3\right)$ | -0.2099 | -0.1541 | -0.1331 |
| $\delta\rho^{Bonding-electron}\left(\vec{r_{ij}}\right)\left(e/Å^3\right)$ | 0.1258 | 0.1314 | 0.07918 |
| $\delta V_{bc}^{bonding}\left(\vec{r_{ij}}\right)(\mathrm{eV})$ | -0.4851 | -1.2864 | -0.7692 |
| $\mu_{A3}$ | -0.5571 | 1.3620 | -1.3683 |

**Table 4** LBS $\varepsilon_x$ (%), relative ACE $\delta E_{\mathrm{C}}$ (%), relative BED $\delta E_{\mathrm{D}}$ (%), indicator of bond nature *m*, BE ratio $\gamma$, overlapping integral *t*, atomic CN $z_x$, FWHM *b*, and BE of Mo, $MoS_2$, $MoSe_2$, and $MoTe_2$.

| | | *b* | $E_v(I)$ | $z_x$ | *t* | $\gamma^m$ | $\gamma$ | *m* | $\varepsilon_x$(%) | $\delta E_{\mathrm{C}}$(%) | $\delta E_{\mathrm{D}}$(%) |
|---|---|---|---|---|---|---|---|---|---|---|---|
| Mo | B | 0.52 | 227.569 | 12.00 | 0.043 | 1.000 | 1 | 1 | 0 | 0 | 0 |

| (100) | $S_1$ | 0.54 | 227.957 | 3.98 | 0.136 | 1.143 | 1.143 | 1 | -12.53 | -62.10 | 70.88 |
|---|---|---|---|---|---|---|---|---|---|---|---|
| | $S_2$ | 0.52 | 227.813 | 5.16 | 0.100 | 1.090 | 1.090 | 1 | -8.27 | -53.13 | 41.23 |
| $MoS_2$ | | 0.99 | 229.210 | 4.00[28] | 0.248 | 1.606 | 1.066 | 3.57 | -12.42 | -46.46 | 139.13 |
| $MoSe_2$ | | 0.91 | 229.282 | 4.00[28] | 0.228 | 1.633 | 1.066 | 3.70 | -12.42 | -45.57 | 143.10 |
| $MoTe_2$ | | 0.50 | 228.380 | 4.00 | 0.125 | 1.300 | 1.066 | 1.98 | -12.42 | -56.68 | 93.48 |